\documentclass[runningheads]{llncs}
\usepackage[most]{tcolorbox}

\newtcolorbox{Tweetquote}[1][]{%
    colback=red!5,
    colframe=red!4,
    notitle,
    enhanced,
    breakable,
    boxsep=0.2mm, left=0.2mm, right=0.2mm, top=0.2mm, bottom=0.2mm,
    }

\usepackage{cite}
\usepackage{algorithm}
\usepackage{multirow}
\usepackage{wrapfig}
\usepackage{amsmath,amsfonts}
\usepackage{algorithmic}
\usepackage{graphicx}
\usepackage{textcomp}
\usepackage{xcolor}
\usepackage{footmisc}
\usepackage{booktabs}
\usepackage{paralist}
\usepackage[autostyle,threshold=0]{csquotes}
\usepackage[hyphens]{url}
\usepackage{url}

\usepackage{hyperref} 
\usepackage[caption=false,font=normalsize,labelfont=sf,textfont=sf]{subfig}
\usepackage{orcidlink}

\usepackage{color}

\hypersetup{
  colorlinks=true,
  linkcolor=blue,
  citecolor=blue,
  urlcolor=blue
}

\begin{document}
\mainmatter 
\title{Duplicating Deceit: Inauthentic Behavior Among Indian Misinformation Duplicators on X/Twitter}
\titlerunning{Misinformation Duplicators} 

\author{
Ashfaq Ali Shafin\,\orcidlink{0000-0002-0135-0091} \and 
Bogdan Carbunar\,\orcidlink{0000-0002-4950-9751}
}

\authorrunning{Shafin and Carbunar} 
\tocauthor{Ashfaq Ali Shafin, Bogdan Carbunar}
\institute{Florida International University, Miami, FL 33199, USA\\
\email{shafinashfaqali21@gmail.com, carbunar@fiu.edu}
}

\maketitle
\vspace{-20pt}
\begin{abstract}
This paper investigates inauthentic duplication on social media, where multiple accounts share identical misinformation tweets. Leveraging a dataset of misinformation verified by AltNews, an Indian fact-checking organization, we analyze over 12 million posts from 5,493 accounts known to have duplicated such content. Contrary to common assumptions that bots are primarily responsible for spreading false information, fewer than 1\% of these accounts exhibit bot-like behavior. We present TweeXster, a framework for detecting and analyzing duplication campaigns, revealing clusters of accounts involved in repeated and sometimes revived dissemination of false or abusive content.
\end{abstract}
\vspace{-10pt}
\section{Introduction}
\vspace{-5pt}

Social media enables rapid and anonymous sharing of information. While these features encourage open communication, they are also often exploited by malicious actors to spread misinformation and abuse~\cite{PLJF23, CDGQS21}. Such manipulation is particularly common during elections and public health crises, when public opinion is most vulnerable~\cite{VNCCGL24}. Coordinated campaigns often use deceptive tactics that include content duplication, to amplify harmful narratives and lend them false credibility. Detecting and understanding these efforts is critical for countering their impact~\cite{TYM22, VRA18}.

To support fact-checking efforts, this paper investigates inauthentic opinion manipulation on X/Twitter, with a focus on identifying candidate content for verification through patterns of duplication. We aim to characterize the actors involved in duplication campaigns and evaluate the predictive value of past behaviors. Specifically, we address the following research questions:

\begin{compactitem}

\item {\bf RQ1}: What kinds of accounts post duplicate misinformation on X/Twitter? What inauthentic behaviors do such accounts exhibit?

\item {\bf RQ2}: Is past involvement in content duplication a predictor of future duplication activities? Can such accounts help identify newly promoted inauthentic content?

\end{compactitem}

We tackle these questions by focusing on post duplication behaviors, where multiple accounts share identical content. To uncover relevant content and associated accounts, we first collected and analyzed misinformation verified by AltNews~\cite{AltNews}, a prominent Indian fact-checking platform. We identified 5,493 accounts that duplicated AltNews-verified misinformation on Twitter prior to its rebranding as X. In 2023, we collected over 12 million posts from these accounts. Our findings reveal that fewer than 1\% misinformation-duplicator accounts exhibit bot-like behavior~\cite{Botometer}, while approximately 4\% are Twitter-verified.

We introduce TweeXster, a framework to study posting activities and analyze content duplication campaigns. TweeXster uncovered inauthentic behaviors including (1) clusters of accounts frequently duplicating content, (2) content repetition and misinformation revival, and (3) campaigns promoting toxic content and specious news sources. Past duplication involvement predicts future activities for both misinformation and abusive speech. Key contributions:
\begin{compactitem}
\item Demonstrate majority of misinformation-disseminating accounts are controlled by real users, not bots [$\S$~\ref{sec:account:analysis}].
\item Introduce TweeXster framework for identifying and studying content duplication campaigns [$\S$~\ref{sec:behaviors:identification}].
\item Provide evidence that AltNews misinformation duplicators persist in duplicating behaviors and inauthentic content dissemination [$\S$\ref{sec:behaviors:inauthentic}, $\S$\ref{sec:content:inauthentic}].
\end{compactitem}
\vspace{-10pt}

\section{Related Work}
\label{sec:related}
\vspace{-5pt}
\noindent {\bf Political Use of X/Twitter in India}. The political use of X/Twitter in India has been extensively studied, revealing highly polarized networks driven by partisanship. Masud and Charaborty~\cite{MC23} analyze posts from the 2022 Indian assembly elections, showing how platforms serve as arenas for self-promotion and political critique. This aligns with Neyazi et al.~\cite{NKS16}, who found Twitter networks polarized by party affiliations and campaign engagement, and Dash et al.~\cite{DMSP22}, who demonstrated that polarized influencers receive higher engagement during political crises, highlighting partisanship's crucial role in platform dynamics.

\noindent {\bf Misinformation Campaigns and Bots}. Misinformation campaigns have become a global challenge, with actors using social media to manipulate public opinion and interfere in democratic processes. Research on Russian disinformation efforts~\cite{ZCBDSB20, NHKE17} 
has shown targeting of electoral processes in the US, UK, and Germany, while in India, campaigns often target religious groups and political opponents~\cite{RPP20}. Although previous studies identify malicious social bots as key amplifiers of misinformation~\cite{CPST19, SCVYFM18}
our findings reveal significant misinformation duplicator accounts are operated by real users rather than bots.

\vspace{-10pt}
\section{Data Collection}
\label{sec:data}
\vspace{-5pt}

\noindent {\bf Fact-Checked Misinformation}. We collected misinformation reports verified by AltNews~\cite{AltNews}, a leading Indian fact-checking organization. We collected 433 reports published between April, 2020 and April, 2022, labeled as follows: politics (237), religion (72), news (62), society (44), media (17), and technology (1). 

\noindent {\bf Misinformation Duplicator Accounts}. We extracted 622 tweets directly linked in AltNews reports as examples of misinformation. Using Twitter search functionality, we identified an additional 6,431 tweets duplicating these 433 instances. We collected metadata for 7,053 tweets, posted by 5,747 unique accounts. We collected additional metadata about these accounts, including profile details and follower/friend counts. We excluded accounts that were (1) suspended or deleted, (2) private, or (3) actively debunking the misinformation. After filtering, we retained 5,493 {\it duplicator accounts}.

\noindent {\bf Tweets of Duplicator Accounts}. In February 2023, we used the Twitter API to collect up to the latest 3,200 tweets for each of the 5,493 duplicators, yielding over 12.7 million tweets. After removing retweets, we obtained 6,879,220 original tweets. To reduce noise, we filtered out posts (1) not written in Hindi or English and (2) containing fewer than four unique words~\cite{LPBF24}. The final dataset, which we refer to as {\bf Dup'23}, consists of 5,070,548 tweets.

\noindent {\bf Specious News Websites}. We curated a comprehensive list of 1,166 specious websites, collected from prior work~\cite{HKD24, MAAG20} and Indian fact-checking organizations~\cite{AltNews, FACTLY, DFRAC}, all of which are known to publish fabricated or misleading content, to verify the presence of these websites among the tweets in the Dup'23 dataset.

\section{Duplicator Account Classification}
\label{sec:account:analysis}
We collected user profile data from the duplicator accounts prior to Elon Musk's acquisition of Twitter. Using the BotometerX API~\cite{Botometer}, we identified 44 duplicator accounts as bots, with both their Complete Automation Probability (CAP) and Raw Bot Score (RBS) exceeding 0.9, to prioritize precision and minimize false positives. The CAP score represents the probability that an account with a given RBS score or higher is automated. Thus, only 0.8\% of the 5,493 active accounts exhibited bot-like behavior. Additionally, 213 accounts were verified by Twitter before the introduction of subscription-based verification on X. Our analysis excludes accounts verified through X’s paid subscription service. Notably, all Twitter-verified accounts had CAP and RBS scores below 0.9. Consequently, the dataset includes 5,236 {\it regular} misinformation duplicator accounts, neither bots nor verified. We analyzed political discourse by leveraging the NivaDuck database \cite{PGAMMCP20} to identify Indian political accounts.

\section{TweeXster Framework}
\label{sec:behaviors}

\subsection{Identifying Clusters of Similar Posts}
\label{sec:behaviors:identification}
To determine if accounts that duplicated AltNews-reported misinformation continue duplicating other content, we identify duplicate tweets in Dup'23 ($\S$~\ref{sec:data}). Since duplicate tweets are not always identical (containing different URLs or other changes), we describe our process for identifying clusters of similar tweets.

\noindent
{\bf Tweet Embeddings}.
We preprocessed tweet text (removing URLs, mentions, etc.), then used SBert with siamese and triplet network architectures to generate embeddings for Dup'23 tweets with more than 4 unique words to reduce noise~\cite{LPBF24}. We employed the paraphrase multilingual MiniLM-L12-v2 transformer model~\cite{NI20} supporting 50+ languages (including English and Hindi). SBert produces 384-dimensional vectors with values between -1 and 1, yielding minimum Euclidean distance of 0 and maximum distance of $\approx 39.19$.

\noindent
{\bf Clustering Tweet Embeddings}.
We used DBSCAN clustering with maximum Euclidean distance of 1, chosen for its ability to identify arbitrary-shaped clusters without pre-specifying cluster count, which is valuable for detecting duplicate tweets where campaign numbers are unknown. To determine the distance threshold, we applied DBSCAN to 7,053 manually identified duplicate misinformation tweets from AltNews, varying distance from 0.1 to 2. A distance threshold of 1 produced 427 clusters, closely aligning with the 433 misinformation reports from AltNews. Of 7,053 misinformation tweets, 7,044 were correctly labeled with only 9 misclassified due to identical tweets in different contexts. Density-based algorithms without explicit distance thresholds (e.g., HDBSCAN) proved less effective, producing clusters with substantial non-duplicate content by grouping tweets based on semantic similarity rather than the strict textual duplication required for identifying exact or near-identical duplicates.

In Dup'23, we identified 172,589 clusters containing 514,958 tweets from 4,938 accounts, with 3,357 clusters having $\geq$10 duplicates and the largest containing 398 duplicates. Clusters averaged 2.98 tweets (SD=3.01). While many clusters appear small, this reflects our limited account subset. Manual searches of 100 randomly-selected small campaigns found $>$25 duplicates per campaign still available on X/Twitter from accounts outside our dataset, suggesting these campaigns represent only the tip of the iceberg of content duplication activities.

To verify cluster consistency, we calculated the cosine similarity between the tweet embeddings within the identified clusters. The lowest observed cosine similarity across all campaigns is 0.91, and the average similarity is 0.99. This indicates a high degree of similarity across all identified clusters, confirming that the campaigns contain near-identical text. Furthermore, we manually analyzed 200 randomly selected clusters to assess their consistency and found that all clusters consisted of duplicate posts, with no anomalies observed.

\begin{wraptable}{r}{0.5\textwidth}
\vspace{-10pt}
\caption{Comparison of duplicate tweets detected by ROPM and our approach.}
\label{tab:wrap_dup_table}
\footnotesize
\centering
\begin{tabular}{lccc}
\toprule
\textbf{Method} & \textbf{\#Tweets} & \textbf{\#Pairs} & \textbf{\#Accs} \\
\midrule
ROPM-10 & 42,820 & 25,934 & 2,613 \\
ROPM-100 & 172,635 & 147,555 & 4,348 \\
Ours & 514,958 & 1,295,785 & 4,938 \\
\bottomrule
\end{tabular}
\vspace{-10pt}
\end{wraptable}

Table~\ref{tab:wrap_dup_table} compares TweeXster with two Ratcliff/Obershelp pattern matching (ROPM)\cite{RM88} baselines: ROPM-10 and ROPM-100. ROPM sorts tweets chronologically and compares each to the next 10 or 100 tweets. ROPM-10 is commonly used for coordinated inauthentic behavior (CIB) detection\cite{VNCCGL24, PFM20}. ROPM-10 and ROPM-100 identify 42,820 and 172,635 duplicate tweets across 2,613 and 4,348 accounts, respectively. TweeXster identifies 514,958 duplicate tweets across 4,938 accounts, yielding 1,295,785 tweet pairs and 736,147 account pairs. These results demonstrate TweeXster captures broader and denser duplication networks than ROPM-based methods, suggesting improved sensitivity for detecting coordinated behaviors.

\subsection{Inauthentic Duplication Behaviors}
\label{sec:behaviors:inauthentic}

We identified 8,118 pairs of accounts that duplicated at least 10 tweets. These pairs involved 1,366 unique accounts, which we refer to as \textbf{\textit{super-duplicator}}. Only three accounts were flagged as bots, while 86 were verified by Twitter. Furthermore, 163 accounts had political affiliations~\cite{PGAMMCP20}, which categorizes accounts based on public declarations and interaction patterns. 79 political accounts were affiliated with the Bharatiya Janata Party (BJP) and 73 with the Indian National Congress (INC), reflecting the two dominant political parties in India.

To understand the structure of duplication behavior, we applied the Louvain community detection algorithm~\cite{BGLL08} to the graph formed by these account pairs. In this graph, nodes represent accounts and edges indicate that the connected accounts duplicated at least 10 tweets in common. The algorithm identified 62 communities with a modularity score of 0.74, suggesting strong intra-community ties and sparse inter-community connections.

\begin{wrapfigure}{r}{0.55\textwidth}
\vspace{-10pt} 
  \centering
  \includegraphics[width=0.57\textwidth]{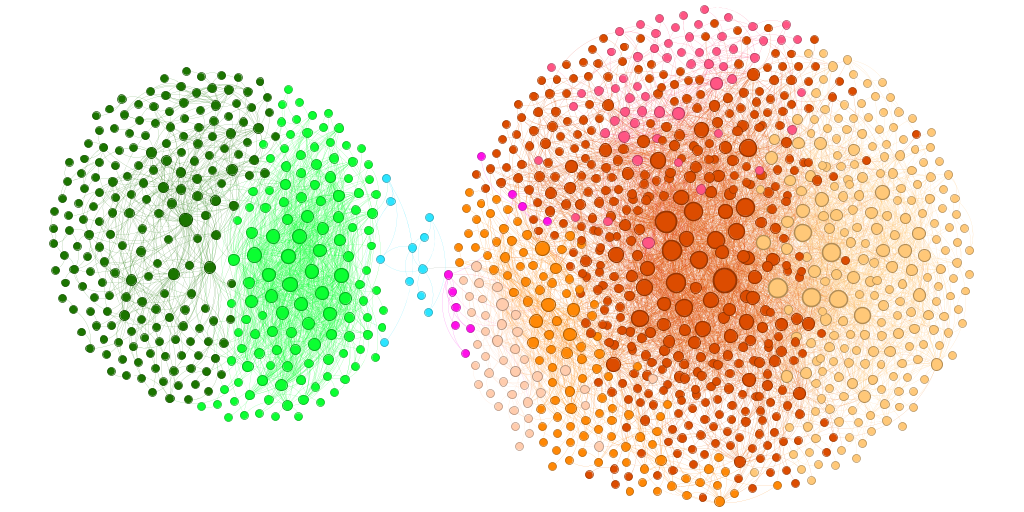}
  \caption{\small Communities of accounts duplicating content. Nodes represent users connected by edges if they duplicated at least 10 posts in common. Node sizes indicate connection degrees. 1,169 accounts (85.6\% of 1,366 total) belong to only 9 clusters.}
  \label{fig:band:cluster}
\vspace{-10pt}
\end{wrapfigure}

The majority of the accounts (85.6\%) belonged to the top nine communities, each containing at least 10 accounts. Figure~\ref{fig:band:cluster} shows these communities, where node size reflects the degree (number of duplication partners). Based on annotations from NivaDuck, six of the largest communities predominantly included BJP-affiliated accounts (shades of red), while two others were largely composed of INC-affiliated accounts (shades of green). One community (blue) did not display any clear political affiliation. Importantly, no single community contained a mix of opposing political affiliations, suggesting alignment in duplication behavior within ideological groups.

\subsection{Patterns of Inauthentic Content}
\label{sec:content:inauthentic}

In the following, we examine the inauthentic methods employed by misinformation duplicators, focusing on patterns of manipulation observed across content dissemination and platform use.

\noindent
{\bf Detecting New Misinformation}. We compiled 20 keywords indicating controversial subjects (e.g., Russia-Ukraine War, Trump, Putin, Biden, Rahul Gandhi, Narendra Modi, Kashmir files, boycott, Adani, Hindutva) or abusive speech against religious minorities/political parties (e.g., bulldozer, stone pelters, love jihad). In the Dup'23 dataset, 79,703 tweets from 32,900 clusters contained at least one keyword. We selected the five longest-active clusters for each keyword, then two researchers independently fact-checked tweets from these 200 clusters using Alt News methodology~\cite{AltNewsMethodology} (Cohen's Kappa k=0.62, indicating moderate agreement). We identified 53 clusters containing misinformation unreported by AltNews and 34 additional clusters with abusive speech. The misinformation clusters contained 209 tweets with 128,658 likes and 65,998 retweets, including false claims about a Pfizer VP arrest, US troops discarding medals, UN Kashmir status changes, and Israeli death sentences for rapists. An example of a tweet containing false information identified through manual fact-checking:

\begin{Tweetquote}
\textit{``VP of Pfizer arrested after leaked documents show only 12\% vaccine efficacy and severe side effects. 
Thanking our govt Pfizer was not allowed in India''}
\end{Tweetquote}

\begin{wraptable}{r}{0.55\textwidth}
\vspace{-20pt}
\caption{Breakdown of toxic duplicate clusters across labels and political parties.}
\label{tab:perspective_result}
\footnotesize
\centering
\begin{tabular}{lccc}
\toprule
\textbf{Label} & \textbf{\# Clusters} & \textbf{\# BJP} & \textbf{\# INC} \\
\midrule
Toxicity         & 4,013 & 2,782 & 639 \\
Severe Toxicity  & 1,088 & 840   & 107 \\
Identity Attack  & 4,226 & 3,276 & 453 \\
Threat           & 1,454 & 1,072 & 179 \\
Insult           & 5,391 & 3,800 & 776 \\
Profanity        &   931 &   642 & 173 \\
\bottomrule
\end{tabular}
\vspace{-15pt}
\end{wraptable}
\vspace{10pt}

\noindent
{\bf Duplicated Toxic Content}. We analyzed the tweet content using Google's Perspective API~\cite{Perspective}, which provides probabilistic scores across multiple toxicity dimensions including identity attacks, threats, profanity, insults, and general toxicity. Tweets receiving scores above 0.5 on any dimension were classified accordingly. Since all the posts in a cluster are near duplicates, we used the Perspective API to retrieve scores for a randomly selected post from each cluster in the Dup'23 ($\S$~\ref{sec:behaviors:identification}). Table~\ref{tab:perspective_result} shows the numbers of clusters with toxicity values over 0.5, including the number of clusters associated with BJP and INC accounts. An example toxic tweet identified by the Perspective API:

\begin{Tweetquote}
\textit{``CHINA must be dragged into International Court and stripped of its VETO power in the UN. `Crime against humanity'. COVID-19 is a Chinese Virus. ''}
\end{Tweetquote}

\begin{wrapfigure}{r}{0.59\textwidth}
\vspace{-10pt}
\centering
\includegraphics[width=0.58\textwidth, height=2cm]{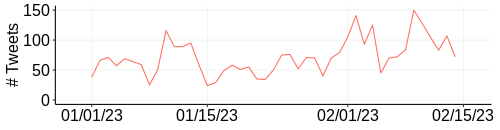}
\vspace{-5pt}
\caption{Timeline of posting of tweets containing links to specious websites during January 1, 2023 to February 15, 2023.} 
\vspace{-10pt}
\label{fig:unreliable_website}
\end{wrapfigure}

\noindent
{\bf Specious News Websites}. In the Dup'23 set, 41,421 tweets contained at least one URL pointing to a specious website (see $\S$~\ref{sec:data}), and 4,249 duplication clusters have tweets with such links. Of the 1,366 super-duplicator accounts, 531 posted at least one tweet with a URL to a specious site, in the Dup'23 dataset, and were responsible for posting 12,822 of the tweets linking to such sites. Accounts affiliated with BJP communities posted 12,440 tweets whereas INC-affiliated accounts only posted 343 tweets. Figure~\ref{fig:unreliable_website} shows the timeline of the number of tweets posted by the super-duplicator accounts, that contain links to specious websites, between January 1 and February 15, 2023. During the first six weeks of 2023, these accounts posted a total of 3,245 tweets linking to a specious site, averaging over 72 tweets per day.

\section{Conclusions}
We analyzed over five thousand X/Twitter accounts duplicating misinformation verified by AltNews using our TweeXster framework, detecting over half a million duplicate tweets and identifying clusters of coordinated inauthentic behavior including sustained misinformation campaigns, toxic discourse, and amplification via specious websites. Remarkably, fewer than 1\% of duplicators were bots, with verified and politically affiliated accounts serving as primary content originators. Our analysis demonstrates that content duplication behavior is both persistent and predictive: accounts sharing past misinformation exhibit significantly higher likelihood of promoting future false content. These findings suggest leveraging historical duplication patterns could enhance fact-checking prioritization, platform moderation, and early campaign detection, offering practical applications for combating online misinformation ecosystems.

\subsubsection{\ackname}

This work was supported in part through NSF awards 2321649 and 2114911. We thank Dr. Lotzi B{\"o}l{\"o}ni for early discussions.

\bibliographystyle{splncs04}
\bibliography{bot,campaigns.bib,cognitive.bib,factcheck.bib,fakenews.bib,influencer.bib,metrics.bib,politicalspam.bib,tweetembedding.bib,deepmodel.bib,hatespeech.bib, tools.bib}

\end{document}